# Neuro-Fuzzy Algorithmic (NFA) Models
# and Tools for Estimation

Danny Ho, Luiz F. Capretz*, Xishi Huang, Jing Ren
NFA Estimation Inc., London, Canada
*University of Western Ontario, London, Canada

## Abstract

Accurate estimation such as cost estimation, quality estimation and risk analysis is a major issue in management. We propose a patent pending soft computing framework to tackle this challenging problem. Our generic framework is independent of the nature and type of estimation. It consists of neural network, fuzzy logic, and an algorithmic estimation model. We made use of the Constructive Cost Model (COCOMO), Analysis of Variance (ANOVA), and Function Point Analysis as the algorithmic models and validated the accuracy of the Neuro-Fuzzy Algorithmic (NFA) Model in software cost estimation using industrial project data. Our model produces more accurate estimation than using an algorithmic model alone. We also discuss the prototypes of our tools that implement the NFA Model. We conclude with our roadmap and direction to enrich the model in tackling different estimation challenges.

## 1. Introducing the NFA Model

The soft computing framework, or NFA Model as presented in Figure 1, consists of the following components:

- Pre-Processing Neuro-Fuzzy Inference System (PNFIS) used to resolve the effect of dependencies among contributing factors of the estimation problem, and to produce adjusted rating values for these factors,
- Neuro-Fuzzy Bank (NFB) used to calibrate the contributing factors by mapping the adjusted rating values for these factors to generate their corresponding numerical parameter values,
- Module that applies an algorithmic model relevant to the nature of the estimation problem to produce one or more output metrics.

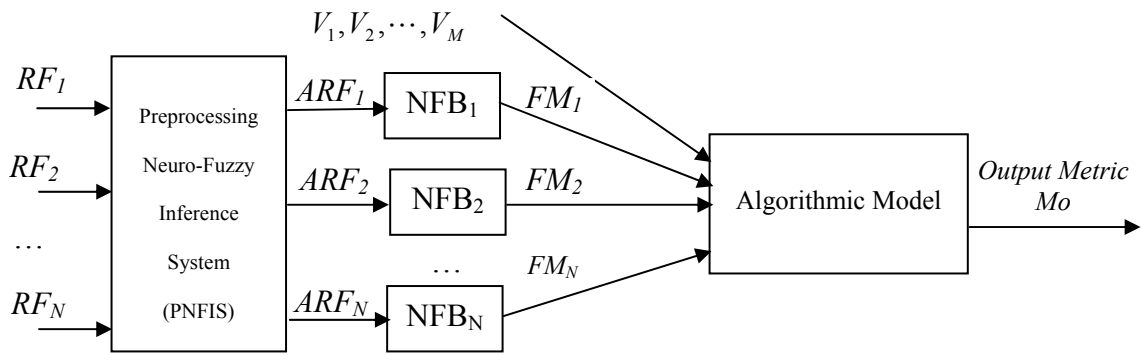

where  N is the number of contributing factors,
M is the number of other variables in the Algorithmic Model,
RF is Factor Rating,
ARF is Adjusted Factor Rating,
NFB is the Neuro-Fuzzy Bank,
FM is Numerical Factor/Multiplier for input to the Algorithmic Model,
V is input to the Algorithmic Model,
and Mo is Output Metric.

**Figure 1**. Neuro-Fuzzy Algorithmic (NFA) Model for Estimation

The NFA Model provides a novel and inventive method for estimation that makes improved use of both historical project data and available expert knowledge, by uniquely combining certain aspects of relatively newer estimation techniques (e.g., neural networks [1] and fuzzy logic [2]) with certain aspects of more conventional estimation models (e.g., algorithmic models such as COCOMO [3] and Function Point Analysis [4]), to produce more accurate estimation results. One big advantage is that the architecture is inherently independent of the choice of algorithmic models and nature of the estimation problems. Our model has learning and adaptation ability, integrates the capability of expert knowledge, project data and parametric algorithmic models, and provides robustness to imprecise and uncertain inputs. It also has good interpretability and high accuracy. As far as we know, this is the first time in software engineering to perform estimation by combining neuro-fuzzy technique with an algorithmic model.

The structure of each element of the Neuro-Fuzzy Bank ($NFB_i$) is shown in Figure 2.

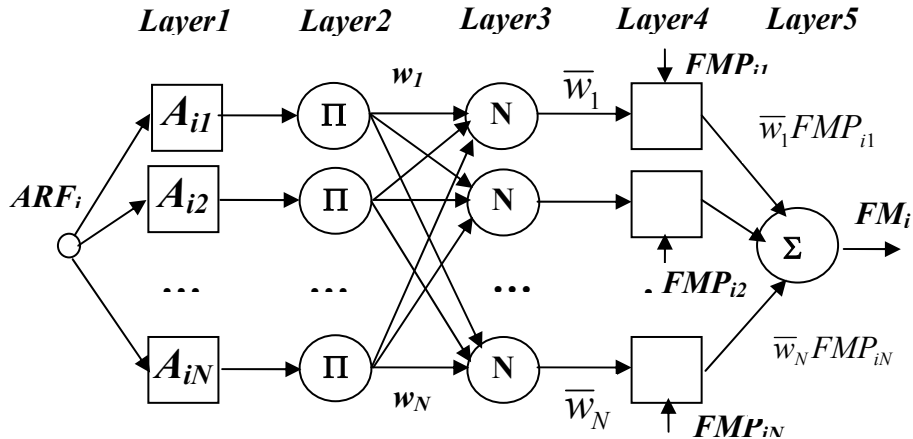

where    $ARF_i$ is Adjusted Factor Rating for contributing factor $i$,

$A_{ik}$ is fuzzy set for the $k$-th rating level of contributing factor $i$,

$w_k$ is firing strength of fuzzy rule $k$,

$\overline{w}_k$ is normalized firing strength of fuzzy rule $k$,

$FMP_{ik}$ is parameter value for the $k$-th rating level of contributing factor $i$,

and    $FM_i$ is numerical value for contributing factor $i$.

**Figure 2**. Elements of $NFB_i$ in the Neuro-Fuzzy Bank

Each layer of the NFB is described as follows:

- Layer 1 is used to calculate the membership values for each fuzzy rule.
- Layer 2 is used to calculate the firing strength for each rule which is the same as the membership value obtained from Layer 1.
- Layer 3 is used to normalize the firing strength for each fuzzy rule.
- Layer 4 is used to calculate the reasoning result of a rule.
- Layer 5 is used to sum up all the reasoning results of fuzzy rules from Layer 4.

## 2.  Validation of the NFA Model

We validated accuracy of software cost estimation for the NFA Model with the following experiments:
- Standard COCOMO Model using 69 project data points [5]
- Stepwise ANOVA Model using 63 project data points [6]
- Function Point Analysis using 184 project data points [7]

Table 1 shows the result of comparison with the COCOMO Model using 69 project data, including six from the industry. The NFA Model improves estimation accuracy when compared with the COCOMO Model alone.

**Table 1**. Estimation Accuracy of NFA Versus COCOMO

| Estimate within Actual (%) | COCOMO Model | | NFA Model | | Improvement |
|---|---|---|---|---|---|
| | # Projects | Accuracy | # Projects | Accuracy | |
| 20% | 49 | 71% | 62 | 89% | 18% |
| 30% | 56 | 81% | 64 | 92% | 11% |
| 50% | 65 | 94% | 67 | 97% | 3% |
| 100% | 69 | 100% | 69 | 100% | 0% |

Table 2 shows the result of comparison with Maxwell and Forselius [8] using the ANOVA approach to build multi-variable models and predict software cost based on 63 project data. The NFA Model improves estimation accuracy when compared with the ANOVA Model, even though the improvement is not very high due to low quality of the dataset with many outliers.

**Table 2**. Estimation Accuracy of NFA Versus ANOVA

| Estimate within Actual (%) | ANOVA | | NFA | | Improvement |
|---|---|---|---|---|---|
| | # Projects | Accuracy | # Projects | Accuracy | |
| 20% | 21 | 33% | 28 | 44% | 11% |
| 25% | 25 | 39% | 33 | 52% | 13% |
| 30% | 33 | 52% | 37 | 58% | 6% |
| 50% | 50 | 79% | 52 | 82% | 3% |
| 100% | 63 | 100% | 63 | 100% | 0% |

Table 3 shows five experimental results of comparison with Function Point Analysis using 184 project data from the International Software Benchmarking Standards Group (ISBSG). The assessment is based on Mean Magnitude Relative Error (MMRE) for estimation accuracy. The average improvement of five experiments is 22%.

**Table 3.** Estimation Accuracy of NFA Versus Function Point Analysis

| | Exp.1 | Exp.2 | Exp.3 | Exp.4 | Exp.5 |
|---|---|---|---|---|---|
| **MMRE Function Point** | 1.38 | 1.58 | 1.57 | 1.39 | 1.42 |
| **MMRE NFA** | 1.10 | 1.28 | 1.17 | 1.03 | 1.11 |
| **Improvement (%)** | 20% | 19% | 25% | 26% | 22% |

We will continue our research effort in validating improvement of the NFA Model over other well-known algorithmic models (e.g., SLIM [9]), and over models in other estimation areas such as size, defect, quality, system-of-systems, system integrator, and so on.

## 3. NFA Tool

We have also built prototypes on the NF COCOMO and NF Function Point products as part of our initial research effort. The tools or final products will adopt the following generic architecture, as depicted in Figure 3:

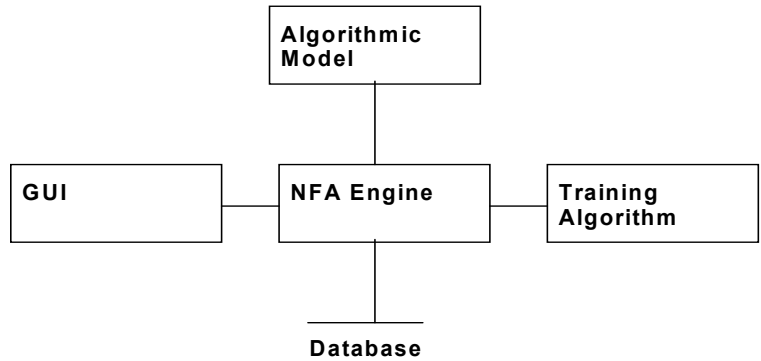

**Figure 3**. Neuro-Fuzzy Algorithmic (NFA) Tool for Estimation

There will be a common front-end GUI, back-end database, and training algorithm. Various types of estimation will be simple plug-in of algorithmic models to the NFA engine, and the effort for additional extension will be small.

- GUI Front-End:
  The GUI front-end has two implementations. The first screen implements the initialization of the PNFIS in Figure 1. Based on a set of system pre-defined defaults, the user is allowed to change the definition of inter-dependencies amongst various contributing factors. The second screen implements the algorithmic model. The user is allowed to input factor ratings and provide other inputs to the model. The output of the GUI will be the set of output metrics of the model as computed by the NFA engine.

- NFA Engine:
  The NFA engine is the brain of the NFA Tool. It implements the NFB as shown in Figure 2. The algorithm is depicted in Figure 4.

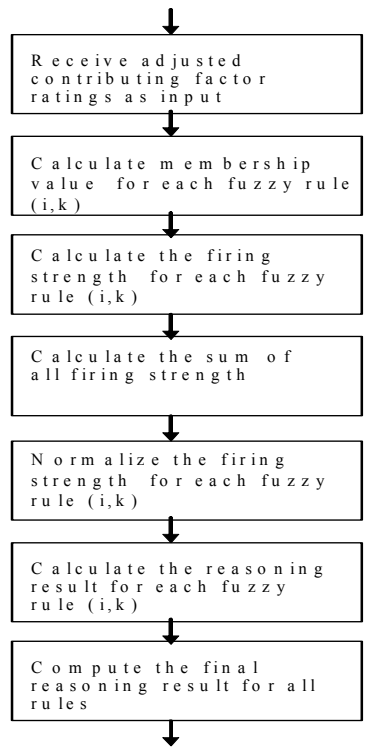

**Figure 4**. Algorithm of the NFA Engine

- Database:
  The database contains valuable project data information for comparing actual with estimation. The purpose of comparison is to train and improve the NFA engine. The project data comes as part of the algorithmic model and grows with local project data captured by the user. Different weights may be assigned for various project data points in training the NFA engine. The database also contains the set of parameter values for different factor ratings. The parameter values may be a set of pre-defined defaults, initialized values from the PNFIS, or values as they mature by training the NFA engine.

- Training Algorithm:
  The training algorithm is key for continuous improvement of the NFA engine. A simple mechanism of MMRE is used to mature the parameter values with project data. The parameter values are subject to monotonic constraint for the factor ratings.

## 4. Future Directions

We believe that our work will help the industry to greatly reduce investment risk by better analyzing the feasibility of projects and effectively managing the development process. Our roadmap follows a two-phase approach for each estimation model or problem: research and commercialization. By treating the NFA Model as a superset of all existing estimation models, the direction of our research is to enrich our soft computing framework to tackle all kinds of estimation challenges. Funding has been secured from the National Sciences and Engineering Research Council (NSERC) of Canada for the next five years.

Our short-term objective is to improve and validate the accuracy of our framework over well-known algorithmic models to handle estimation problems. Some initial software specific estimation areas to tackle include software cost estimation, size estimation, quality estimation, systems of systems estimation, and system integrator estimation, among others. We see opportunities in partnering with existing estimation methodologists in improving the accuracy of existing estimation models. The long term objective is to generically apply this framework to other aspects of estimation such as prediction of stock performance [10], investment risk estimation, prediction of medical condition of patients, disease growth, and so on. This will be accomplished in cross discipline research and development.

We also see opportunities in partnering with existing estimation tool vendors in producing a series of potential new products such as NF COCOMO, NF Function Point, NF SLIM, NF Size, NF Defect, NF SoS, NF SysInteg, NF Stock, NF Medical, and so forth. Lastly, we plan to refine our intellectual property by looking into bundling the NFA Model with other estimation techniques. We also need to take advantage of the latest development in neural network, fuzzy logic and algorithmic models, and enrich the NFA Model in accurately and consistently providing estimation results.